\documentclass[prl,aps,twocolumn,amsmath,amssymb,longbibliography]{revtex4-1}

\usepackage{multirow}
\usepackage{amssymb}
\usepackage{amsmath}
\usepackage{mathrsfs}
\usepackage{graphicx}
\usepackage{amsfonts}
\usepackage{color}
\usepackage{bm}
\usepackage{xspace,soul}

\newcommand{\sqrtm}[1]{\sqrt{\smash[b]{m_{#1}}}}

\begin{document}
\title{An exactly solvable quantum four-body problem associated with the symmetries of an octacube}

\author{Maxim Olshanii}
\email{maxim.olchanyi@umb.edu}
\affiliation{Department of Physics, University of Massachusetts Boston, Boston Massachusetts 02125, USA}
\author{Steven G.\ Jackson}
\affiliation{Department of Mathematics, University of Massachusetts Boston, Boston Massachusetts 02125, USA}

\date{\today}


%
\begin{abstract}
In this article, we show that eigenenergies and eigenstates of a system consisting of four one-dimensional
hard-core particles with masses $6m$, $2m$, $m$, and $3m$ in a hard-wall box
can be found exactly using Bethe Ansatz. The Ansatz is based on the exceptional affine reflection group  $\tilde{F}_{4}$
associated with the symmetries and tiling properties of an {\it octacube}---a Platonic solid unique to four dimensions,
with no three-dimensional analogues.
We also uncover the Liouville integrability structure of our problem: the four integrals
of motion in involution are identified as invariant polynomials of the finite reflection group $F_{4}$, taken as functions of the components of
{\it momenta}.
\end{abstract}

\maketitle

\textit{Introduction}.--
The relationship between exactly solvable quantum one-dimensional multi-particle problems and kaleidoscopes has been long appreciated
\cite{gaudin1971_386}.
A kaleidoscope is a system of mirrors where none of the multiple images of the original object is broken at the mirror junctions. A viewer situated inside a kaleidoscopic cavity
has no means of distinguishing the images of objects from the original.
For a broad class of boundary conditions,
the eigenmodes of a kaleidoscopic cavity are represented by {\it finite} superpositions of plane waves; they can be found {\it exactly}, using the method of images \cite{sutherland1980_1770}. When a many-body problem reduces to a solvable kaleidoscope,
the resulting solution is known as the (coordinate) Bethe Ansatz solution \cite{gaudin1983_book,sutherland2004_book}. The list of particle systems solvable using Bethe Ansatz is so far exhausted by:
equal mass hard-core bosons, on an open line, on a circle or in between walls \cite{girardeau1960_516}; $\delta$-interacting, equal-mass, generally distinguishable particles on a line and on a circle \cite{mcguire1963_622,yang1967_1312,gaudin1967_0375}; this includes bosons \cite{lieb1963_1605}, which could also be in the presence of one or two walls \cite{gaudin1971_386}; systems of two hard-core particles with masses $m$ and
$3m$ interacting with a wall, for both wall-$m$-$3m$ and wall-$3m$-$m$ spatial orders were
briefly commented on  in \cite{mcguire1963_622}, but discarded as inferior to the problems with finite strength interactions.

It has been long conjectured that no exceptional---specific to a given number of spatial dimensions---kaleidoscopes lead to solvable
particle problems with realistic interactions \cite{gaudin1971_386,gaudin1983_book}. The search for physical realizations was limited
to systems of particles of {\it the same mass}, with a possible addition of immobile walls.
In this Letter, we show that
the exceptional closed kaleidoscope  $\tilde{F}_{4}$ induces a novel quantum integrable system:
four one-dimensional quantum hard-core particles with {\it different} masses, $6m$, $2m$, $m$, and $3m$, in a hard-wall box.
The solution utilizes the symmetries of an octacube, a Platonic solid unique to four dimensions.

\textit{Identifying the particle problem generated by the Coxeter diagram $\tilde{F}_{4}$}.---
Consider six hard-core (i.e.\ impenetrable) particles on a line, with masses $m_{0}$, $m_{1}$, $m_{2}$, $m_{3}$, $m_{4}$, and $m_{5}$. Their
coordinates will be denoted as $x_{0}$, $x_{1}$, $x_{2}$, $x_{3}$, $x_{4}$, and $x_{5}$ respectively.
The natural coordinate transformation
\begin{align*}
x_{i} = \sqrt{\frac{{\cal M}}{m_{i}}}\, y_{i}\,\, \mbox{, for } i=0,\,1,\,\ldots,\, 5
%
\end{align*}
reduces the system to a single six-dimensional particle of mass ${\cal M}$ (to be fixed later) whose motion is constrained by the following set of five inequalities:
$
y_{i+1}/\sqrtm{i+1} > y_{i}/\sqrtm{i}$, for
$
i=0,\,1,\,\ldots,\, 4
$.

\begin{figure}
\centering\includegraphics[width=.45\textwidth]{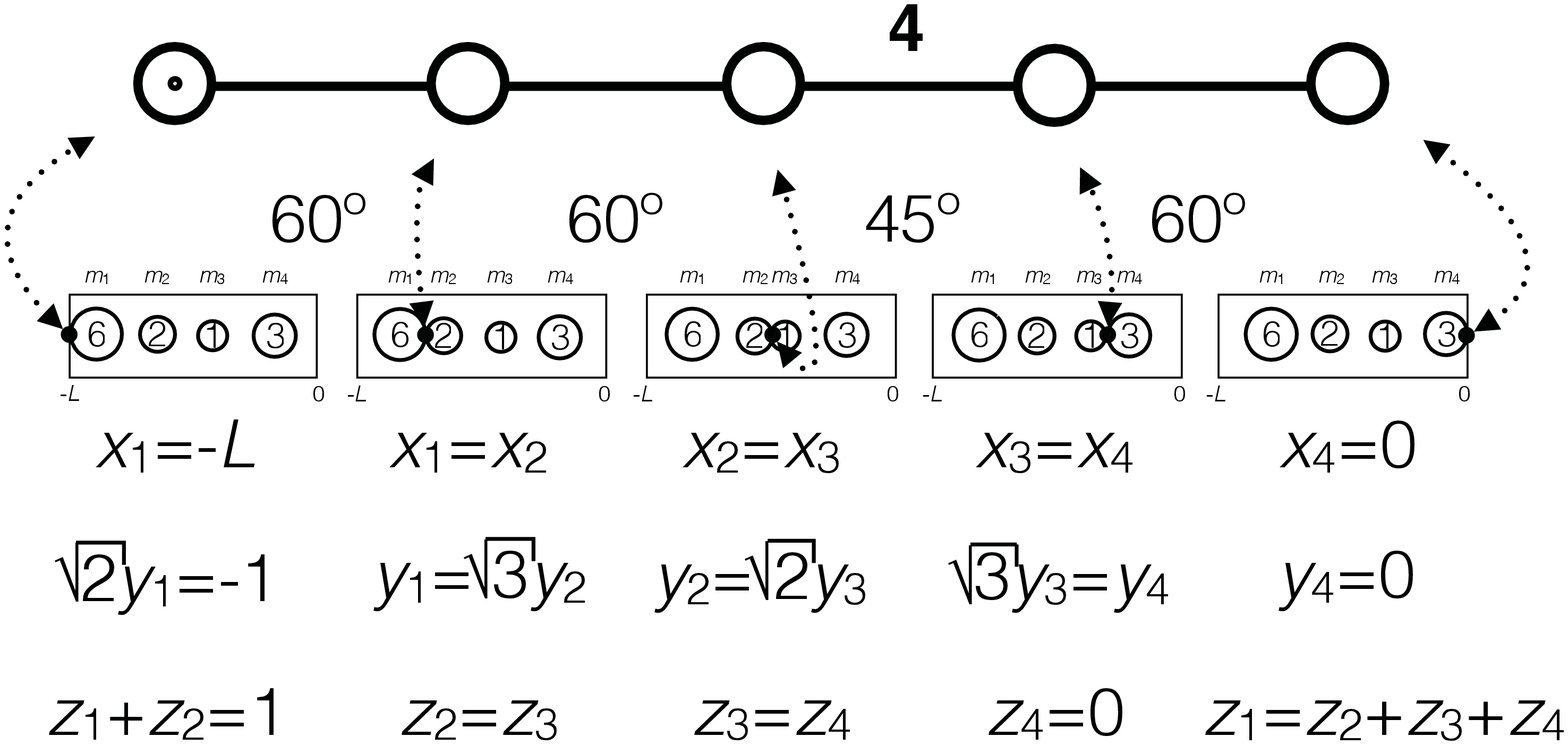}
\caption{
The $\tilde{F}_{4}$ kaleidoscope and its particle realization. From top to bottom: the affine Coxeter diagram $\tilde{F}_{4}$; the angles between the
mirrors;  the four particles with masses $6m$, $2m$, $m$, and $3m$ between two hard walls at $x_{0}=-L$ and $x_{5} = 0$;
the particle-particle and particle-wall hyperplanes of contact in all three coordinate systems used in the main text.
}
\label{f:coxeter_diagram_workv}
\end{figure}

Consider now the affine Coxeter diagram $\tilde{F}_{4}$  \cite{coxeter_book_1969} depicted in the top line of Fig.~\ref{f:coxeter_diagram_workv}.
It encodes the geometry of a particular simplex-shaped 4-dimensional kaleidoscope. The vertices label
its five mirrors:  the 3-faces (represented by 3-dimensional simplexes) of the 4-dimensional simplex. The edges---the links---of the diagram (and their absences) encode the angles between the corresponding mirrors, i.e. the angles between their normals. When two vertices are not linked by an edge, the respective mirrors are orthogonal to each other. An edge with no numbers above it corresponds to an angle of $\pi/3$. A number $n$ above an edge would produce an angle of $\pi/n$ between the corresponding mirrors.
The meaning of the mark inside the leftmost vertex will be explained later.
In our case, the non-right angles between the mirrors are $\pi/3$, $\pi/3$, $\pi/4$, and $\pi/3$,
in the order of their appearance on the Coxeter diagram, left to right.

Let us try to identify the five consecutive vertices of the Coxeter diagram
with the five hyperplanes of contact between neighboring particles:
$\frac{y_{i+1}}{\sqrtm{i+1}} = \frac{y_{i}}{\sqrtm{i}}$ for
$i=0,\,1,\,\ldots,\, 4$ (see the particle diagrams in Fig.~\ref{f:coxeter_diagram_workv}). This choice is natural: contact hyperplanes for
two non-overlapping pairs of neighboring particles are indeed orthogonal to each other. Then the four edges of the diagram correspond to the four triplets of consecutive particles.


For three consecutive particles, $i$, $i+1$, and $i+2$, the angle between the $i$ vs $i+1$ and $i+1$ vs $i+2$ contact hyperplanes is given by
\begin{align*}
\theta_{i\,(i+1)\,(i+2)} = \arctan\sqrt{
                                                        \frac{
    m_{i+1}(m_{i}+m_{i+1}+m_{i+2})
                                                               }
                                                               {
   m_{i}m_{i+2}
                                                               }
                                              }
\,\,,
\end{align*}
(see e.g.\ \cite{mcguire1963_622}).
Notice that these angles do not depend on the overall mass scale. Therefore, any constraints on these angles can only lead to relationships between the five
mass ratios, $m_{i+1}/m_{i}$ for $i=0,\,1,\,2,\,3,\,4$ alone, with no overall mass scale involved. On the other hand, for our six-body system to lead  to the $\tilde{F}_{4}$ kaleidoscope,
the angles between the hyperplanes of contact must satisfy four equations only,
\begin{align*}
\theta_{012} = \frac{\pi}{3};\, \theta_{123}= \frac{\pi}{3};\,
\theta_{234} = \frac{\pi}{4};\, \theta_{345} = \frac{\pi}{3}
\,\,,
\end{align*}
thus being seemingly underdetermined.
Nevertheless, once the non-negativity of the masses ($m_{i} \ge 0$ for $i=0,\,1,\,2,\,3,\,4,\,5$) is invoked,
a single solution (up to an overall scale $m$) survives:
\begin{align*}
&
m_{0} =+\infty;\,m_{1} = 6m;\,m_{1} = 2m;
\\
&
m_{3} = m;\,m_{4} = 3m;\, m_{5}=+\infty
%
%
\,\,.
\end{align*}
In retrospect, that both the leftmost and the rightmost mass should be infinite could have been predicted from the onset: this is
probably the only way to map six particles on an {\it open} line to a single 4-dimensional particle in a {\it closed} cavity.

Without loss of generality, we can place the infinitely massive particles at
\begin{align*}
&
x_{0} = -L;\, x_{5} =0
\,\,,
\end{align*}
where $L$ is the size of the resulting hard-wall box (see the hyperplanes of contact in terms of the original $x$-coordinates in  Fig.~\ref{f:coxeter_diagram_workv}).
Now observe that $m_{3}=m$ and $L$ appear to be convenient mass and length scales, and $\hbar$ a natural scale of action. From now on, we will be using
the system of units where
\begin{align*}
m_{3} = L = \hbar = 1
\,\,.
\end{align*}
Also, it will become clear from what follows that the choice ${\cal M} = 2 m_{1}$ allows us to respect the existing conventions
on the size of the octacube.

Figure~\ref{f:coxeter_diagram_workv} also shows the corresponding formulae for the contact hyperplanes expressed through the transformed coordinates $y_{1}$, $y_{2}$, $y_{3}$, and $y_{4}$.
These hyperplanes form the boundaries of the $\tilde{F}_{4}$ kaleidoscope we have in mind:
$
-1 <\sqrt{2}\, y_{1}
$ ;
$
y_{1} < \sqrt{3}\,y_{2}
$;
$y_{2} < \sqrt{2}\, y_{3}$
;
$
\sqrt{3}\,y_{3} < y_{4}
$
;
$\,y_{4}<0$.

The coordinate transformation
${\bm z} = \hat{T}_{{\bm z} \gets {\bm y}}\cdot{\bm y}$, with the rotation-inversion matrix $\hat{T}_{{\bm z} \gets {\bm y}}$ given by
\[
\hat{T}_{{\bm z} \gets {\bm y}}=\frac{1}{2\sqrt{3}}\left(
\begin{array}{cccc}
 -\sqrt{6} 	& -\sqrt{2} 	& -1 			& -\sqrt{3} \\
 -\sqrt{6} 	& \sqrt{2} 		& 1			& \sqrt{3} \\
  0 		& -2 \sqrt{2}  	& 1 			& \sqrt{3} \\
  0		& 0			& -3	 		& \sqrt{3}
\end{array}
\right)\,,
\]
brings the domain of our $\tilde{F}_{4}$ kaleidoscope to
the conventional form \cite{gaudin1983_book,gutkin1979_6057,sutherland1980_1770,gutkin1982_1,emsiz2006_191,emsiz2009_571,emsiz2010_61}:
\begin{align}
\begin{split}
&
{\cal D}_{\hat{e}} \equiv 
\Big\{
{\bm z}\mbox{ such that }\,
{\bm \alpha}_{0}\cdot{\bm z} >  -\frac{1}{2} {\bm \alpha}_{0}\cdot{\bm \alpha}_{0} = -1
\\
&
\qquad\qquad
\mbox{ and }
{\bm \alpha}_{1,\,2,\,3,\,4}\cdot{\bm z} > 0
\Big\} 
\,\,,
\end{split}
\label{4D_simplex_z}
\end{align}
where ${\bm y} \equiv (y_{1},\,y_{2},\,y_{3},\,y_{4})$, and ${\bm z} \equiv (z_{1},\,z_{2},\,z_{3},\,z_{4})$.

According to this convention, the kaleidoscope (\ref{4D_simplex_z}) is a 4-dimensional simplex with
five 3-faces (mirrors) defined by five inward normals ${\bm \eta}_{j} = {\bm \alpha}_{j}/|{\bm \alpha}_{j}|$ for $j=0,\,1,\,2,\,3,\,4$,
with
$
{\bm \alpha}_{0} = (-1,\,-1,\,0,\,0)
$,
$
{\bm \alpha}_{1} = (0,\,+1,\,-1,\,0)
$,
$
{\bm \alpha}_{2} = (0,\,0,\,+1,\,-1)
$,
$
{\bm \alpha}_{3} = (0,\,0,\,0,\,+1)
$,
$
{\bm \alpha}_{4} = (+\frac{1}{2},\,-\frac{1}{2},\,-\frac{1}{2},\,-\frac{1}{2})
$
being the so-called minimal (or negative of maximal) root ($j=0$)
and the simple roots ($j=1,\,2,\,3,\,4$) of the corresponding finite reflection group $F_{4}$ \cite{humphreys_book_1997}.
The mirrors from 1 to 4 pass through the origin; the 0th mirror (marked by a dot in the Coxeter diagram of Fig.~\ref{f:coxeter_diagram_workv}) passes through the point $(1/2,\,1/2,\,0,\,0)$.
The hyperplanes that define the inequalities (\ref{4D_simplex_z})  are also identified in Fig.~\ref{f:coxeter_diagram_workv}.
In Eq.~\ref{4D_simplex_z}, the subscript the subscript $\hat{e}$ in ${\cal K}_{\hat{e}}$
has the meaning of the identity element of the finite group $F_{4}$, a convention whose meaning will become clear 
from what follows.
Note that the ``natural coordinates'' ${\bm z}$ can be expressed through the particle coordinates ${\bm x}$ as
${\bm z} = \hat{T}_{{\bm z} \gets {\bm x}}\cdot{\bm x}$, with
\[
\hat{T}_{{\bm z} \gets {\bm x}}=\frac{1}{12}\left(
\begin{array}{cccc}
 -6 		& -2 		& -1 			& -3 \\
 -6 		&  2 		& 1			& 3 \\
  0 		& -4 		& 1 			& 3 \\
  0		& 0		& -3	 		& 3
\end{array}
\right)\,.
\]

\textit{Periodic tiling by consecuitive reflections}.---
The key to Bethe Ansatz solvability of the closed kaleidoscopic billiards lies in their ability to
 {\it periodically} tile the full space---with neither holes nor overlaps---via consecutive reflections about their faces
\cite{sutherland1980_1770,distinct_faces}.
Consider the $\tilde{F}_{4}$ kaleidoscope (\ref{4D_simplex_z}).
A union of the figures produced by sequential  reflections of  (\ref{4D_simplex_z}) about its 3-faces from the 1-st to the 4-th (``simple root'' mirrors),
\begin{align}
\begin{split}
{\cal C} &\equiv \cup_{\hat{g} \in F_{4}} {\cal D}_{\hat{g}}
\\
&=  
\Big\{
{\bm z}\mbox{ such that }\,
(\hat{g}{\bm \alpha}_{0})\cdot{\bm z} >  -\frac{1}{2} (\hat{g}{\bm \alpha}_{0})\cdot(\hat{g}{\bm \alpha}_{0}) = -1,
\\
&\qquad
\mbox{ for all  }\hat{g} \in F_{4}
\Big\}
\,\,,
\end{split}
\label{4D_cell_z}
\end{align}
with
\begin{align}
\begin{split}
&
{\cal D}_{\hat{g}} \equiv 
\Big\{
{\bm z}\mbox{ such that }\,
(\hat{g}{\bm \alpha}_{0})\cdot{\bm z} >  -\frac{1}{2} (\hat{g}{\bm \alpha}_{0})\cdot(\hat{g}{\bm \alpha}_{0}) = -1
\\
&
\qquad\qquad
\mbox{ and }
(\hat{g}{\bm \alpha}_{1,\,2,\,3,\,4})\cdot{\bm z} > 0
\Big\}
\,\,,
\end{split}
\label{4D_simplex_z__ALL}
\end{align}
leads to the so-called octacube,
otherwise known as the 24-cell \cite{coxeter_book_1969}; it is the only 4-dimensional Platonic solid that
does not have any 3-dimensional analogues.  Its $24$ vertices lie at points given by all coordinate permutations and sign choices applied to
$(\pm 1,\,0,\,0,\,0)$ and $(\pm \frac{1}{2},\,\pm \frac{1}{2},\,\pm \frac{1}{2},\,\pm \frac{1}{2})$. Its $24$ octahedron-shaped 3-faces are centered
at all coordinate permutations and sign choices of $(\pm \frac{1}{2},\,\pm \frac{1}{2},\, 0,\, 0)$.
Sequential reflections about the ``simple root'' mirrors of the $\tilde{F}_{4}$ kaleidoscope
form the symmetry group of the octacube, the reflection group $F_{4}$. (The same label
is used to mark both an open kaleidoscope, where the $0$th, ``minimal root'' mirror is absent and the corresponding simple, ``non-affine'' Coxeter
diagram.)
The union in (\ref{4D_cell_z}) runs over all elements $\hat{g}$ of the group $F_{4}$. The ``physical'' domain
(\ref{4D_simplex_z}) is the identity member of the set (\ref{4D_simplex_z__ALL}).

The 4-dimensional space can be periodically tiled by identical octacubes; the centers of the octacubical cells
lie on all points with integer coordinates whose coordinates sum to an even number:
$
\left\{
  (z_{1},\,z_{2},\,z_{3},\,z_{4}):\,\,z_{1}+z_{2}+z_{3}+z_{4} = \mbox{even}
\right\}
$.
The very same tiling can be obtained by a consecutive reflection of the (\ref{4D_simplex_z}) simplex about all five of its mirrors, including the ``minimal root'' one.
Transformations generated by these reflections form the symmetry group of the octacubical tiling---
the affine reflection group $\tilde{F}_{4}$, homonymous to the corresponding closed kaleidoscope.

In order to facilitate the possible classification of energy levels by symmetry, we choose the lattice vectors ${\bm a}_{1,2,3,4}$ to
be proportional to
the four simple roots of the reflection group $D_{4}$ \cite{humphreys_book_1997},
closely associated with $F_{4}$:
${\bm a}_{1} = (1,\,-1,\,0,\,0)$, ${\bm a}_{2} = (0,\,1,\,-1,\,0)$, ${\bm a}_{3} = (0,\,0,\,1,\,-1)$, and 
${\bm a}_{4} = (0,\,0,\,1,\,1)$ . The
reciprocal lattice vectors are thus ${\bm \kappa}_{1} = (2\pi,\,0,\,0,\,0)$, ${\bm \kappa}_{2} = (2\pi,\,2\pi,\,0,\,0)$, ${\bm \kappa}_{3} = (\pi,\,\pi,\,\pi,\,-\pi)$, and ${\bm \kappa}_{4} = (\pi,\,\pi,\,\pi,\,\pi)$.

\textit{Finding the eigenenergies and eigenstates of the problem}.---
In general, finding the eigenenergies and eigenstates of a $D$-dimensional kaleidoscopic billiard amounts to solving a system of $D$
nonlinear algebraic equations---the so-called Bethe Ansatz equations \cite{gaudin1983_book,gutkin1979_6057,sutherland1980_1770,gutkin1982_1,emsiz2006_191,emsiz2009_571,emsiz2010_61}. However,
in the case of hard-core interactions, the problem greatly simplifies (see Corollary 3.2 in \cite{emsiz2006_191}).  The eigenstates acquire the form
\begin{align}
\psi({\bm z}) = \frac{1}{\sqrt{V_{{\cal D}} |G|}} \sum_{\hat{g}\in G} (-1)^{{\cal P}(\hat{g})}  \exp[i (\hat{g}{\bm k}) {\bm z}]
\,\,,
\label{generic_eigenstate}
\end{align}
where $\hat{g}$ are the members of the finite reflection group $G$($=F_{4}$ in our case) generated by sequential reflections about the ``simple root'' mirrors of the kaleidscope in question;
${\cal P}(\hat{g})$ is the parity of the number of the elementary reflections needed to reach $\hat{g}$ from the identity transformation;
$V_{{\cal D}} = V_{{\cal C}}/|G|$($ =1/576$ in our case) is the kaleidoscope volume; 
$V_{{\cal C}}=|\det[({\bm a}_{1},\,{\bm a}_{2},\,{\bm a}_{3},\,\ldots)]|$ is the 
unit cell volume;
$|G|$($=1152$ for $F_{4}$) is the order, i.e.\ the number of elements, of the group $G$.
The conjecture contained in Theorem 1 in \cite{emsiz2010_61} indicates that the wavefunction (\ref{generic_eigenstate}) is normalized to unity:
$\int_{{\cal D}_{\hat{e}}} dz_{1} dz_{2} dz_{3}\ldots  |\psi({\bm z})|^2 = 1$ when integrated over the corresponding kaleidoscope 
${\cal D}_{\hat{e}}$
((\ref{4D_simplex_z}) in the case of $\tilde{F}_{4}$). This conjecture covers not only the case of 
the hard-wall boundary conditions but also a much more general class of boundary conditions, associated with finite strength interactions. 
The wavevectors $k$ are simply drawn from the reciprocal lattice of the corresponding tiling \cite{open_simplexes_general}.
In order to prevent  both double counting and the formal appearance of the eigenstates that are identically zero,
the wavevectors $k$ must be further restricted to those lying inside the open kaleidoscope associated with the kaleidoscope in question, obtained
by removing the ``minimal root'' mirror: ${\bm \eta}_{j}\cdot {\bm k} > 0$ for $j=1,\,2,\,3,\,\ldots$ (see the end of  Sec.~3 of \cite{emsiz2006_191}).

For the case of the hard-wall boundary conditions, the normalization formula used above can also be proven directly 
without resorting to the conjecture \cite{emsiz2010_61}. Observe that the octacube (\ref{4D_cell_z}) is a unit cell
of the lattice spanned by the lattice vectors ${\bm a}_{1,\,2,\,3,\,4}$ and that the ``seed'' wavevector ${\bm k}$, along with 
all its images $\hat{g}{\bm k}$ belong to the respective reciprocal lattice. In that case, any two distinct plane waves 
with wave vectors
$\hat{g}{\bm k}$  and $\hat{g}'{\bm k}$ will be orthogonal to each other if integrated over the 
whole cell:
\begin{align*}
\langle \hat{g}'{\bm k} | \hat{g}{\bm k} \rangle_{{\cal C}} = 
\delta_{\hat{g}',\,\hat{g}} V_{{\cal C}}
\,\,,
\end{align*}
with
\begin{align*}
\langle {\bm k}_{1} | {\bm k}_{2} \rangle_{{\cal A}} 
\equiv
\int_{{\cal A}} \!d^{d}{\bm z} \exp[i ({\bm k}_{2} - {\bm k}_{1}){\bm z}]
\,\,,
\end{align*}
where ${\cal A}$ is an area of space. The normalization integral in question can now be evaluated as
\begin{align*}
&
\int_{{\cal D}_{\hat{e}}} \!d^{d}{\bm z} |\sum_{\hat{g}\in G} (-1)^{{\cal P}(\hat{g})}  \exp[i (\hat{g}{\bm k}) {\bm z}]|^2 
\\
&\quad
= 
\sum_{\hat{g}'} \sum_{\hat{g}} (-1)^{{\cal P}(\hat{g}')+{\cal P}(\hat{g})} 
\langle \hat{g}'{\bm k} | \hat{g}{\bm k} \rangle_{{\cal D}_{\hat{e}}}
\\
&\quad
\stackrel{\hat{g}''\equiv \hat{g}^{-1}\hat{g}'}{=}
\sum_{\hat{g}''} \sum_{\hat{g}} (-1)^{{\cal P}(\hat{g}\hat{g}'')+{\cal P}(\hat{g})} 
\langle \hat{g} \hat{g}''{\bm k} | \hat{g}{\bm k} \rangle_{{\cal D}_{\hat{e}}} 
\\
&\quad
=
\sum_{\hat{g}''} (-1)^{{\cal P}(\hat{g}'')} 
\sum_{\hat{g}} 
\langle \hat{g} \hat{g}''{\bm k} | \hat{g}{\bm k} \rangle_{{\cal D}_{\hat{e}}}  
\\
&\quad
\stackrel{\hat{g}'''\equiv \hat{g}^{-1}}{=}
\sum_{\hat{g}''} (-1)^{{\cal P}(\hat{g}'')} 
\sum_{\hat{g}'''} 
\langle \hat{g}''{\bm k} | {\bm k} \rangle_{{\cal D}_{\hat{g}'''}}  
\\
&\quad
=
\sum_{\hat{g}''} (-1)^{{\cal P}(\hat{g}'')}  
\langle \hat{g}''{\bm k} | {\bm k} \rangle_{{\cal C}}  
=
\sum_{\hat{g}''} (-1)^{{\cal P}(\hat{g}'')}  
\delta_{\hat{g}',\,\hat{e}} V_{{\cal C}}
\\
&\quad
= V_{{\cal C}} = V_{{\cal D}} |G|
\,\,;
\end{align*}
this result justifies the choice of the normalization factor in the expression  (\ref{generic_eigenstate}).
Above, we used $(-1)^{{\cal P}(\hat{g}\hat{g}'')} = (-1)^{{\cal P}(\hat{g})} (-1)^{{\cal P}(\hat{g}'')}$ and 
$(-1)^{2{\cal P}(\hat{g})}=1$.

We are now in the position to write down the spectrum of our system explicitly. Starting from this point
we abandon the  $m_{3} = L = \hbar = 1$ system of units.
The eigenenergies, $E = (\hbar^2/2{\cal M}) k^2$, and the eigenstates (\ref{generic_eigenstate}),
with ${\bm k} = \sum_{j=1,2,3,4} n_{j} {\bm \kappa}_{j}$, $n_{j=1,2,3,4} \in \mathbb{Z}$, and ${\bm \eta}_{j=1,2,3,4}\cdot {\bm k} > 0$,
are then given by
\begin{align}
&
E_{n_{1}n_{2}n_{3}n_{4}} = \frac{\pi^2\hbar^2}{6 m_{3} L^2}
\left(
   2n_{2}(n_{1}+n_{2}+n_{3}+n_{4})
   +
\right.
\label{spectrum}
\\
&
\qquad
\left.
   n_{1}^2+n_{3}^2+n_{4}^2
   +
   n_{1}n_{3}+n_{1}n_{4}+n_{3}n_{4}
\right)
\nonumber
\\
&
\Psi_{n_{1}n_{2}n_{3}n_{4}} ({\bm x})
\label{eigenstates}
\\
&
\qquad
= \frac{1}{\sqrt{48 L^4}}
 \sum_{\hat{g}} (-1)^{{\cal P}(\hat{g})}  \exp[i (\hat{g} {\bm k}_{n_{1}n_{2}n_{3}n_{4}} ) \cdot \hat{T}_{{\bm z} \gets {\bm x}} \cdot {\bm x} ]
\nonumber
\\
&
\mbox{for }
n_{2} \ge 1
\mbox{ and }
n_{1} > n_{4} > n_{3} \ge 1
\nonumber
\,,
\end{align}
where
\begin{align*}
{\bm k}_{n_{1}n_{2}n_{3}n_{4}} = \left(
                                                      \begin{array}{c}
                                                       2 n_{1} + 2 n_{2} + n_{3} + n_{4}
                                                       \\
                                                       2 n_{2} + n_{3} + n_{4}
                                                       \\
                                                       n_{3} + n_{4}
                                                       \\
                                                       -n_{3} + n_{4}
                                                      \end{array}
                                                    \right) \, \frac{\pi}{L}
\,\,;
\end{align*}
the sum in (\ref{eigenstates}) runs over all $1152$ members $\hat{g}$ of the reflection group $F_{4}$ \cite{coxeter_book_regular_polytopes}
(transformations generated by consecutive reflections about the four ``simple root'' mirrors of the $\tilde{F}_{4}$ kaleidoscope);
$\Psi_{n_{1}n_{2}n_{3}n_{4}} ({\bm x}) \equiv \sqrt{|\det[\hat{T}_{{\bm z} \gets {\bm x}}]|} \,
\psi_{n_{1}n_{2}n_{3}n_{4}} (\hat{T}_{{\bm z} \gets {\bm x}} \cdot  {\bm x}) $
are the eigenstates of the system expressed through the particle coordinates $x_{1},\, x_{2},\,x_{3},\,x_{4}$,
normalized as
$\int_{x_{1}=-L}^{0} \int_{x_{2}=x_{1}}^{0} \int_{x_{3}=x_{2}}^{0} \int_{x_{4}=x_{3}}^{0} d^{4}{\bm x} |\Psi_{n_{1}n_{2}n_{3}n_{4}}({\bm x})|^2 = 1$;
$\left( \frac{{\cal D} {\bm z}}{{\cal D} {\bm x}} \right)$ is the Jacobian matrix
of the ${\bm z} \gets {\bm x}$ transformation.

\begin{figure}
\centering\includegraphics[width=.45\textwidth]{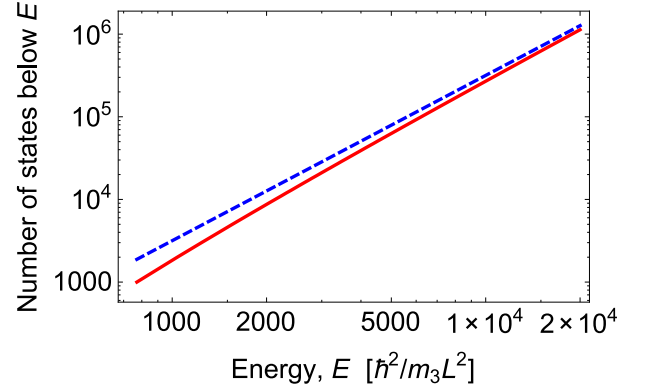}
\caption{(color online). Exact energy spectrum for four particles of mass $m_{1}=6m$, $m_{2}=2m$, $m_{3}=m$, and $m_{4}=3m$ in a hard-wall box of length $L$ (red solid line).
Weyl's law prediction for the spectrum of this system (blue dashed line).The energies up to  $2\times 10^{4} \hbar^2/m_{3}L^2$ are shown.}
\label{f:weyl_s_law__EMax_2E4__SAVE}
\end{figure}

At high energies, the spectrum converges to the Weyl law prediction for the number of states
with energies below a given energy $E$:
$
{\cal N}(E) = \frac{{\cal W}(E)}{(2\pi\hbar)^{D=4}} = \frac{m_{3}^2 L^4 E^2}{32 \pi^2 \hbar^4}
$,
where ${\cal W}(E)$ is the classical phase-space volume occupied by points
with energies below $E$, and $D$ is number of spatial dimensions (see Fig.\ \ref{f:weyl_s_law__EMax_2E4__SAVE}).

\begin{figure}
\centering\includegraphics[width=.45\textwidth]{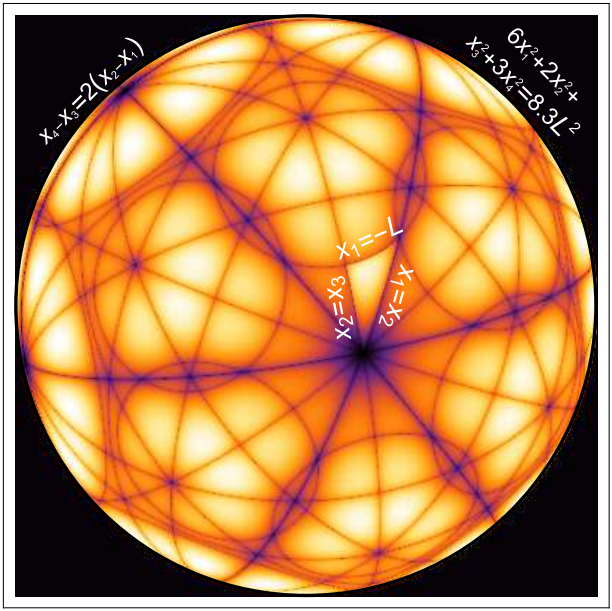}
\caption{(color online). The ground state of our four-body system. The plot is performed using the $z$ coordinates, relevant to the tiling. The image is produced by first intersecting the 4-dimensional density distribution by the 3-plane and the 3-sphere indicated in the upper left and upper right corners, respectively (the formulas, however, are given in $x$ coordinates, in which the 3-sphere becomes a 3-ellipsoid). The result is a 2-sphere centered at the origin.  Considering now this 2-sphere as living in the usual 3D space, we pick, in this 3D space, a (2D) plane that does not intersect the 2-sphere, and project onto it the 2-sphere's hemisphere closest to the plane; this projection is what is shown in the image. The circles on the 2-sphere are the sections of the 3D hyperplanes of the original 4D space; the great circles come from the hyperplanes that pass through the origin, and the other circles come from the hyperplanes that do not.  The lighter triangle in the middle corresponds to a section of the  ``physical'' 4-dimensional simplex
to which the particle coordinates are bounded; the
remainder of the sphere is a section of the entire 4-dimensional space when this space is octacubically tiled with the mirror images of the
original ``physical'' simplex. The existence of this tiling ensures the applicability of the Bethe Ansatz \cite{sutherland1980_1770}.}
\label{f:sphere_0p1m1m1_R1OverSqrt2Plus.125__SAVE_BETTER}
\end{figure}

The ground state energy is
\begin{align}
E_{\mbox{\scriptsize ground state}} = E_{n_{1}=3,\,n_{2}=1,\,n_{3}=1,\,n_{4}=2} = \frac{13 \pi^2\hbar^2}{2 m_{3} L^2}
\,\,.
\label{E_GS}
\end{align}
The ground state wave function can be obtained from Eq.~(\ref{eigenstates}), using
\begin{align}
{\bm k}_{\mbox{\scriptsize ground state}} = {\bm k}_{n_{1}=3,\,n_{2}=1,\,n_{3}=1,\,n_{4}=2} =
\left(
                                                      \begin{array}{c}
                                                       11
                                                       \\
                                                       5
                                                       \\
                                                       3
                                                       \\
                                                       1
                                                      \end{array}
                                                    \right) \, \frac{\pi}{L}
\,\,.
\label{k_GS}
\end{align}
Fig.\ \ref{f:sphere_0p1m1m1_R1OverSqrt2Plus.125__SAVE_BETTER}
shows a particular section of the ground state density distribution within
the $\tilde{F}_{4}$ simplex, along with its space-tiling mirror images.

\textit{Integrals of motion}.---
The Bethe Ansatz integrability of our system can be also reinterpreted in terms of the Liouville integrability.
To construct the three additional integrals of motion in involution with the Hamiltonian and each other,
we suggest invoking the invariant polynomials of the group in question \cite{chevalley1955_778}: finite polynomials of coordinates $w(z_{1},\,z_{2},\,z_{3},\,\ldots)$ that remain
invariant under the group action,
\begin{align}
w(\hat{g} {\bm z}) = w({\bm z})
\,\,.
\label{invariant_polynomials}
\end{align}
Consider now an operator $\hat{I}$ that is constructed by taking an invariant polynomial $w$ as a function of the \textit{momenta} associated with the ${\bm z}$ coordinates:
\begin{align}
\hat{I} \equiv w(-i {\bm \nabla_{{\bm z}}})
\,\,.
\nonumber
\end{align}
Let us show that any eigenstate (\ref{generic_eigenstate}) of the kaleidoscope associated with the group in question is at the same time an eigenstate of $\hat{I}$. Indeed,
\begin{eqnarray*}
&&
\hat{I} \psi({\bm z}) =
\\
&&
\quad w(-i {\bm \nabla_{{\bm z}}})  \sum_{\hat{g}} (-1)^{{\cal P}(\hat{g})}  \exp[i (\hat{g}{\bm k}) {\bm z}] =
\\
&&
\quad\quad  \sum_{\hat{g}} (-1)^{{\cal P}(\hat{g})}  w(\hat{g}{\bm k}) \exp[i (\hat{g}{\bm k}) {\bm z}]  =
\\
&&
\quad\quad\quad  w({\bm k})  \sum_{\hat{g}} (-1)^{{\cal P}(\hat{g})}  \exp[i (\hat{g}{\bm k}) {\bm z}]
\,\,,
\end{eqnarray*}
Q.E.D. For the reflection group $F_{4}$, the four lowest power functionally independent invariant polynomials read  \cite{mehta1988_1083}:
\begin{eqnarray}
&&
\!\!\!\!\!\!\!\!\!\!\!\!\!\!\!\!
\begin{split}
w_{M}^{(F_{4})}({\bm z})  =
(z_{1}-z_{2})^{l_{M}} +
(z_{1}+z_{2})^{l_{M}} +
(z_{1}-z_{3})^{l_{M}} +
\\
(z_{1}+z_{3})^{l_{M}} +
(z_{1}-z_{4})^{l_{M}} +
(z_{1}+z_{4})^{l_{M}} +
\\
(z_{2}-z_{3})^{l_{M}} +
(z_{2}+z_{3})^{l_{M}} +
(z_{2}-z_{4})^{l_{M}} +
\\
(z_{2}+z_{4})^{l_{M}} +
(z_{3}-z_{4})^{l_{M}} +
(z_{3}+z_{4})^{l_{M}}
\end{split}
\label{invariant_polynomials_F4}
\\
&&
\!\!\!\!\!\!\!\!\!\!\!\!\!\!\!\!
l_{M=1}=2;\,l_{M=2}=6;\,l_{M=3}=8;\,l_{M=4}=12
\nonumber
\,\,.
\end{eqnarray}
Accordingly, the four fundamental integrals of motion for 
our system read
\begin{eqnarray}
&&
\hat{I}_{M}^{(F_{4})} =w_{M}^{(F_{4})}(
     ((\hat{T}_{{\bm z} \gets {\bm x}})^{-1})^{\top} \cdot \hat{{\bm p}}_{{\bm x}}
                                                       )
\label{integrals_of_motion}
\\
&&
M=1,\,2,\,3,\,4
\,\,,
\nonumber
\end{eqnarray}
with
\begin{eqnarray}
\hat{{\bm p}}_{{\bm x}} \equiv
(-i\hbar\frac{\partial}{\partial x_{1}}, \, -i\hbar\frac{\partial}{\partial x_{2}}, \, -i\hbar\frac{\partial}{\partial x_{3}}, \, -i\hbar\frac{\partial}{\partial x_{4} })
\end{eqnarray}
being the particle momenta, corresponding to the particle coordinates
$x_{1}$, $x_{2}$, $x_{3}$, $x_{4}$. Above, $(\ldots)^{\top}$ stands for the transpose of a matrix. Note that the first integral
of motion is proportional to the Hamiltonian of the system: $\hat{I}_{1}^{(F_{4})}/m = 144 \hat{H}$.

\textit{Summary and outlook}.--
In this Letter, we obtain---using Bethe Ansatz---an exact expression for the eigenenergies and eigenfunctions of four hard-core particles
with mass ratios $6\!:\!2\!:\!1\!:\!3$ in a hard-wall box.
The Ansatz is induced by a hidden symmetry of the system related to the
symmetries of the tiling of a 4-dimensional space by octacubes. The exact spectrum stands in good agreement with the approximate Weyl's law prediction.

The following observation may serve as a seed for a longer research program. The procedure, outlined in this Letter, for identifying a few-body problem relevant to a particular Coxeter diagram
\cite{coxeter_book_regular_polytopes} is
not unique to $\tilde{F}_{4}$.
Any diagram, affine or not,
that {\it does not have bifurcations} can potentially be used to generate a solvable few-body hard-core problem.
The $\tilde{A}_{N-1}$
($N$ identical hard-cores on a circle) and $\tilde{C}_{N}$ ($N$ identical hard-cores in a hard-wall box)
diagrams have been already successfully introduced in the first years of many-body Bethe Ansatz, in \cite{girardeau1960_516} and \cite{gaudin1971_386}, respectively.
The $I_{2}(n)$ diagrams were explored in 
Ref.\ \cite{hwang2015_13319}, 
albeit classically, still awaiting a quantum treatment:
for each integer $n$ a continuous one-parametric family of solvable three-body mass ratios (on a line with no boundary conditions) was obtained.
A system of two hard-core particles with mass ratio 3:1 in a box (whose one-wall version was explored in Ref.~\cite{mcguire1963_622})
is expected to exhibit an exact solution associated with the $\tilde{G}_{2}$ Coxeter diagram.
It can be conjectured that $H_{3}$ and $H_{4}$ Coxeter
diagrams lead to solvable four- and five-body problems on a line, respectively.

The true challenge is posed by the Coxeter diagrams (affine or otherwise) {\it with bifurcations}, most notably the $\tilde{E}_{6,\,7,\,8}$ series. At the moment, it is not clear if there are
any realistic many-body problems that can be solved using these symmetries. (Some examples of ``unphysical''
many-body realizations of Coxeter diagrams---e.g. those where a given particle can interact with an empty point in between the
other two---are given in Ref.~\cite{gaudin1983_book}.)

The potential empirical context for this and further planned explorations is one-dimensional cold gas mixtures in optical lattices where the dynamics is governed by an effective
(rather than physical) mass, controlled at will \cite{gadway2011_145306}.

\textit{Acknowledgments}.--
We are profoundly indebted to late Marvin Girardeau for guidance and inspiration.
We are thanking Jean-S\'{e}bastien Caux for reading the manuscript and for the comments that followed.
The impact of the numerous in-depth discussions of the subject
with Dominik Schneble can not be overestimated. This work was supported by grants from National Science Foundation ({\it PHY-1402249}) and
the Office of Naval Research ({\it N00014-12-1-0400}).

\bibliography{Bethe_Ansatz_v010,f4_paper_additional_refs_01}

\begin{thebibliography}{23}
\bibitem [{dis()}]{distinct_faces}%
  \BibitemOpen
  \href@noop {} {}\bibinfo {note} {Regardless of the geometry, all faces are
  assumed to be distinct.}\BibitemShut {Stop}%
\end{thebibliography}

\end{document}